\documentclass[%
 reprint,
 amsmath,amssymb,
 aps,
prb]{revtex4-1}

\usepackage{subfigure}
\usepackage{graphicx}
\usepackage{dcolumn}
\usepackage{bm}

\begin{document}

\preprint{APS/123-QED}

\title{Long-range Coulomb interaction effects on the surface Dirac electron system of a three-dimensional topological insulator}

\author{Nobuyuki Okuma}
\author{Masao Ogata}%
\affiliation{%
 Department of Physics, The University of Tokyo, Hongo 7-3-1, Tokyo 113-0033, Japan
}%

%
\date{\today}

\begin{abstract}
The surface state of a three-dimensional topological insulator forms a two-dimensional massless Dirac electron system. In Dirac electron systems, Coulomb interaction is not screened due to the small density of states at the Fermi energy and thus the long-range Coulomb interaction (LRCI) plays a crucial role. In this paper, 
we investigate the surface state with chemical potential $\mu=0$ in the presence of the LRCI using the Wilsonian renormalization group.
We first check the Fermi velocity enhancement in the surface Dirac system, which also occurs in a usual Dirac electron system. The most remarkable feature of the surface Dirac system is that the Dirac Hamiltonian contains not pseudo spin but real spin Pauli matrices. 
Because of this feature, we find the g-factor enhancement, which is a unique property of the surface Dirac system.
We also investigate the explicit form of the spin susceptibility and find that the spin susceptibility is enhanced in the presence of the LRCI.

\begin{description}
\item[PACS numbers]
75.70.-i,\ 71.10.-w,\ 73.20.At
\end{description}

\end{abstract}

\maketitle


\section{introduction}
The physics of topological insulators is one of the most important subjects in recent condensed matter physics$\cite{1, 2}$.
A topological phase realized in a topological insulator is characterized by a non-trivial band structure, which has a non-zero $Z_2$ or $Z$ topological invariant.
Although the bulk of a topological insulator is a band insulator, its boundary state is a topologically protected metal.
Several models of realistic topological insulators with strong spin-orbit interaction$\cite{3, 4, 5}$ have been proposed theoretically, and some of them have been realized experimentally$\cite{6, 7}$. 
Topological phases have been generalized to gapped materials like superconductors and the periodic table of topological phases in any spatial dimension$\cite{8,9}$ has been obtained.

One of the reasons why topological insulators attract a lot of interest is their theoretical simplicity due to the one-body nature of the band theory. 
Since the fundamental classification of non-interacting topological insulators has been completed, one of the remaining major issues is many-body effects in these materials.
Some works have focused on phase transitions in topological insulators in the presence of short-range interactions$\cite{10}$.
For instance, the Kane-Mele model with Hubbard interaction has been investigated by both analytical$\cite{11}$ and numerical$\cite{12}$ calculations. 
Other works targeted topological phases driven not by spin-orbit interactions but by short-range interactions$\cite{13}$.
However, the screening effect is generally weak in insulators, semiconductors, and semimetals due to the small density of states at the Fermi energy.
Therefore it is a non-trivial problem whether or not we can assume a short-range interaction for a topological insulator.
In particular, the surface state of a three-dimensional topological insulator like Bi$_2$Se$_3$ forms a massless Dirac electron system$\cite{14}$ like graphene.
It is well known that the long-range Coulomb interaction (LRCI) plays a crucial role in a usual Dirac electron system$\cite{15}$.
We should therefore assume the LRCI at least in the surface state.

Effects of the LRCI have been well investigated in a massless Dirac electron system.
Some works have investigated interaction effects in graphene using the Wilsonian renormalization group analysis$\cite{16, 17, 18, 19}$.
These works have shown that the Fermi velocity is a relevant coupling constant in the 1-loop level renormalization group analysis.
This velocity enhancement would also occur in the surface state, which is a new type of massless Dirac electron system.

However, there is a fundamental difference between graphene and the surface Dirac electron system. 
Pauli matrices in the effective Hamiltonian for graphene represent pseudo spin. On the other hand, Pauli matrices in the surface Dirac electron system represent real spin.
This would cause a different spin response to the magnetic field. 

In this paper, we investigate the spin response of the surface state of a three-dimensional topological insulator in the presence of the LRCI using the Wilsonian renormalization group analysis. 
We find a renormalization of the g-factor which is absent in graphene. Also we show that the spin susceptibility is enhanced compared with the non-interacting case.
In sect.2, we introduce a model of the surface Dirac system and discuss some justification of perturbation theory. In sect.3, we introduce the Wilsonian renormalization group analysis at 1-loop level and check the Fermi velocity renormalization, which occurs in a usual Dirac electron system. In sect.4, we derive the renormalization group equation for the g-factor and show the g-factor renormalization, which is a unique property of the surface state. In sect.5, we first show that the spin susceptibility of the surface state is totally different from a usual Dirac electron system even in the non-interacting case. Then we calculate the spin susceptibility of the surface state in the presence of the LRCI. Although the Fermi velocity renormalization decreases the spin susceptibility, we obtain the enhanced spin susceptibility due to the g-factor renormalization. 

\section{MODEL}
We investigate the linearized surface Dirac Hamiltonian of a three-dimensional topological insulator
\begin{equation}
\mathcal{H}_{sur}(\textbf{k})=v(k_x\sigma_y-k_y\sigma_x),\label{hamiltonian}
\end{equation}
where $v$ is the Fermi velocity and $\sigma$'s represent spin Pauli matrices. We set the momentum cutoff $\Lambda$ and assume the chemical potential $\mu=0$ in the following. The bare thermal Green function is defined as
\begin{equation}
\mathcal{G}(\textbf{k}, i\omega_n)=\frac{1}{i\omega_n-v(k_x\sigma_y-k_y\sigma_x)},
\end{equation}
where $\displaystyle \omega_n=\pi T (2n+1)$ represents the $n$th fermionic Matsubara frequency.
In this paper, we assume the two-dimensional LRCI $\cite{15}$
\begin{equation}
V(\textbf{q})=\frac{2\pi e^2}{\epsilon |\textbf{q}|},
\end{equation}
due to the non-screening nature of Dirac electron systems. 
We treat this interaction as a perturbation.

The magnitude of the interaction compared with the bare Hamiltonian 
\begin{equation}
\alpha=\frac{e^2}{v\epsilon}
\end{equation}
is a constant of order unity in usual Dirac electron systems$\cite{15}$.
It seems that this relatively large $\alpha$ breaks the perturbative renormalization group.
However, previous studies of the perturbative renormalization group in Dirac electron systems $\cite{15,16,17,18,19}$ have shown that the Fermi velocity is renormalized to large values.
This will justify the perturbative renormalization group analysis at low energy scale.
Although the connection between small $\alpha$ region and large one is still unclear, we perform the perturbative renormalization group analysis in this paper as in the previous studies.
\section{Fermi velocity renormalization}
In this section, we perform the 1-loop Wilsonian renormalization group (WRG) analysis of the Hamiltonian ($\ref{hamiltonian}$) in the presence of the LRCI. 
Although the same Fermi velocity renormalization has been obtained in previous studies$\cite{15,16,17,18,19}$, we show the WRG analysis which is used in the later sections.
We treat the partition function
\begin{equation}
Z=\int \mathcal{D}(\bar{\Psi},\Psi)\mathrm{exp}[-S_0-S_{int}], \label{start}
\end{equation}
where
\begin{align}
S_0=&\frac{1}{\beta}\sum_{n}\int^{\Lambda}\frac{d\textbf{k}}{(2\pi)^2}\bar{\Psi}(\textbf{k},i\omega_n)\left[-\mathcal{G}^{-1}(\textbf{k},i\omega_n)\right]\Psi(\textbf{k},i\omega_n),\\
S_{int}&=\frac{1}{2\beta^3}\sum_{l,m,n}\sum_{\sigma,\sigma '}\int^{\Lambda} \frac{d\textbf{k}d\textbf{k}'d\textbf{q}}{(2\pi)^6}\bar{\psi}_{\sigma}(\textbf{k}+\textbf{q},i\omega_l)\notag\\
&\bar{\psi}_{\sigma '}(\textbf{k}'-\textbf{q},i\omega_m)\psi_{\sigma '}(\textbf{k}',i\omega_n)\psi_{\sigma}(\textbf{k},i\omega_{l+m-n})V(\textbf{q}).
\end{align}
Here $\bar{\Psi}=(\bar{\psi}_{\uparrow}, \bar{\psi_{\downarrow}})$ is the spinor field for electrons with up and down spins. 

In the first step of the WRG analysis, we integrate out fast fields $\Psi_>$, whose momentum is in the momentum shell $\delta\Lambda$ ($\Lambda e^{-\delta l}\leq |\textbf{k}|\leq\Lambda$), and obtain the effective action for slow fields $\Psi_<$, whose momentum is inside the momentum sphere 
($|\textbf{k}|\leq\Lambda e^{-\delta l} $). The result at 1-loop level is as follows (See reviews of the WRG in condensed matter \cite{20}):
\begin{equation}
Z\cong Const. \int \mathcal{D}(\bar{\Psi}_<,\Psi_<)\mathrm{exp}\left[-S_0-S_{int}-\delta S\right], \label{Const}
\end{equation}
where
\begin{equation}
\delta S=\frac{1}{\beta}\sum_{n}\int^{\Lambda e^{-\delta l}}\frac{d\textbf{k}}{(2\pi)^2}\bar{\Psi}(\textbf{k},i\omega_n) \delta \Sigma(\textbf{k},i\omega_n)\Psi(\textbf{k},i\omega_n).\label{deltas}
\end{equation}
Usually, we focus on the action to obtain RG equations and neglect the prefactor, like $Const.$ in eq.($\ref{Const}$), of the partition function.
However, in order to study the free energy, we take account of the prefactor (sect.5).

\begin{figure}[htbp]
\begin{center}
　　　\includegraphics[width=3cm,angle=90,clip]{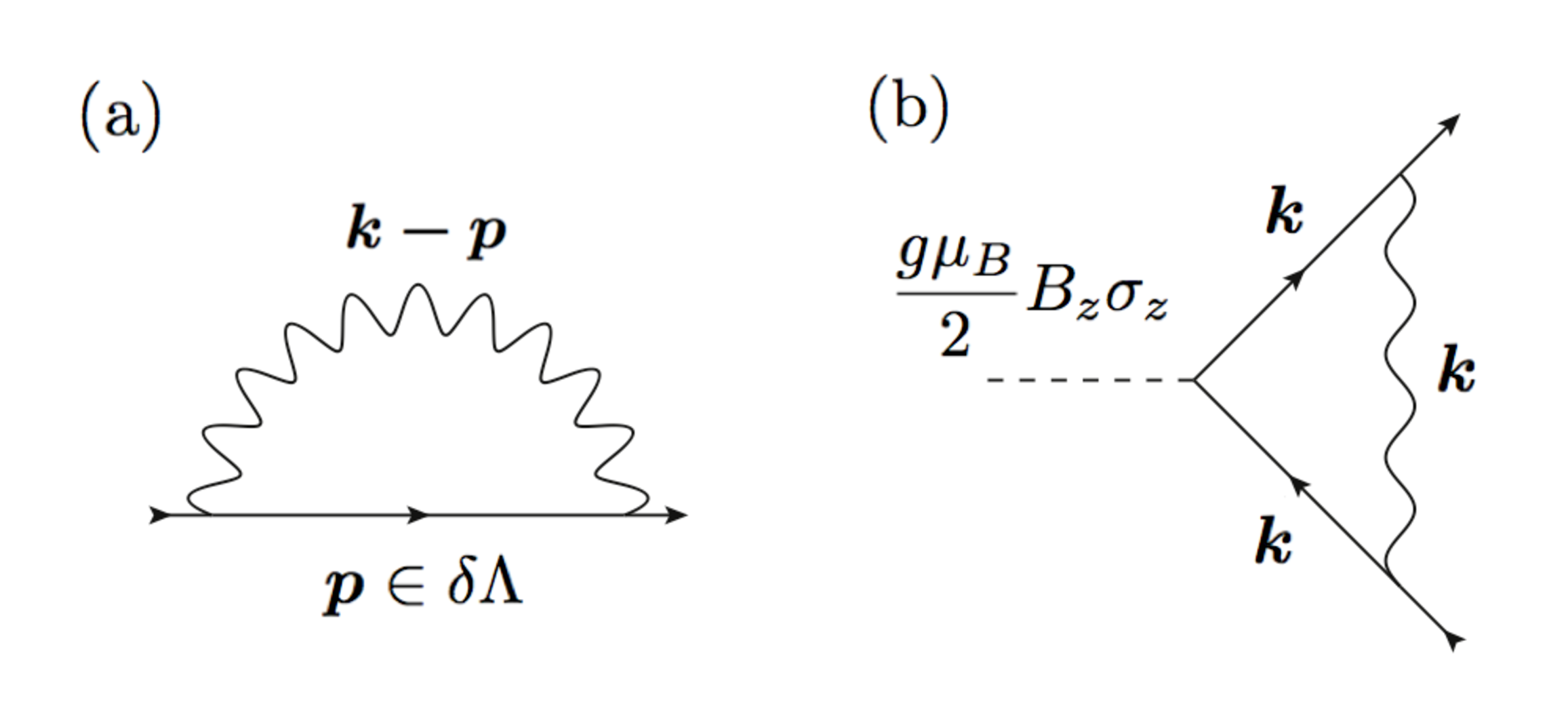}
　　　\caption{Feynman diagrams considered in sect.3 and sect.4 : (a) correction to the kinetic term $\delta\Sigma(\textbf{k},i\omega_n) $, (b) correction to the Zeeman term $\delta\Gamma.$}
　　　\label{fig1}
\end{center}
\end{figure}

$\delta\Sigma(\textbf{k},i\omega_n)$ in eq.($\ref{deltas}$) is represented by Fig. $\ref{fig1}$(a) and the momentum integral is in the momentum shell $\delta\Lambda$.
We can easily compute this diagram as
\begin{align}
\delta\Sigma(\textbf{k},i\omega_n)&=-\frac{1}{\beta}\sum_{\omega_n}\int_{\textbf{p}\in \delta\Lambda}\frac{d\textbf{p}}{(2\pi)^2}\mathcal{G}(\textbf{p},i\omega_n)V(\textbf{k}-\textbf{p})\notag\\
&\cong\delta l\frac{e^2}{4\epsilon}(k_x\sigma_y-k_y\sigma_x).
\end{align} 
As a result, the effective quadratic action $S_0'=S_0+\delta S$ becomes
\begin{align}
S_0'[\bar{\Psi}_<,\Psi_<]=&\frac{1}{\beta}\sum_{n}\int^{\Lambda e^{-\delta l}}\frac{d\textbf{k}}{(2\pi)^2}\bar{\Psi}(\textbf{k},i\omega_n)\biggl[-i\omega_n\textbf{1}\biggr.\notag\\
&\biggl. +\left(v+\frac{e^2}{4\epsilon}\delta l \right)(k_x\sigma_y-k_y\sigma_x)\biggr]\Psi(\textbf{k},i\omega_n),
\end{align}
where $\delta S$ changes the quadratic term.

In the second step of the WRG analysis, momenta are rescaled so that the action has the same momentum cutoff as the original one.
In this rescaling procedure, we also have to rescale frequencies and fields to keep the form of the quadratic action.
In the finite temperature WRG, rescaling frequencies means enhancing the temperature.
After this rescaling, we obtain the effective quadratic action

\begin{align}
S_0'[\bar{\Psi}',\Psi']=&\frac{1}{\beta'}\sum_{n}\int^{\Lambda}\frac{d\textbf{k}'}{(2\pi)^2}\bar{\Psi}'(\textbf{k}',i\omega_n')\biggl[-i\omega_n'\textbf{1}\biggr.\notag\\
&\biggl. +\left(v+\frac{e^2}{4\epsilon}\delta l \right)(k_x'\sigma_y-k_y'\sigma_x)\biggr]\Psi'(\textbf{k}',i\omega_n') \ \ ,
\end{align}
where 
\begin{align}
&\beta'=\beta e^{-\delta l},\omega_n'=\omega_ne^{\delta l}, \textbf{k}'=\textbf{k}e^{\delta l},\notag\\ 
&\Psi'(\textbf{k}', i\omega_n')=e^{-2\delta l}\Psi(\textbf{k}'e^{-\delta l},i\omega_n'e^{-\delta l}).
\end{align}
One can see that the quartic action does not change in this rescaling.

Now we can construct the RG equation for the Fermi velocity:
\begin{equation}
\frac{dv(l)}{dl}=\frac{e^2}{4\epsilon}.
\end{equation}
From this, we obtain the effective Fermi velocity
\begin{equation}
v(l)=v+\frac{e^2}{4\epsilon}l.
\end{equation}
The WRG procedures are performed until the temperature becomes the energy cutoff:
\begin{equation}
Te^{l^*}=v(l^*)\Lambda,\label{tvrelation}
\end{equation}
where the magnitude of the interaction $\alpha$ is sufficiently small and we can approximately regard this system as a non-interacting case. 
In the temperature range of $T\ll T_0\equiv v\Lambda$, $l^*$ can be roughly written as 
\begin{equation}
l^*\approx\mathrm{log}\frac{T_0}{T},
\end{equation} 
and the Fermi velocity can be written in the form
\begin{equation}
v(T)\approx v+\frac{e^2}{4\epsilon}\mathrm{log}\frac{T_0}{T}.\label{velocity}
\end{equation}
Eq.($\ref{velocity}$) shows that the Fermi velocity is enhanced by the LRCI at low temperatures.
This enhancement of the Fermi velocity also occurs in graphene$\cite{15}$.

The difference between the surface Dirac Hamiltonian and graphene Hamiltonian is the meaning of "spins" represented by Pauli matrices.
Pauli matrices in the surface state represent real spins, while those in graphene represent pseudo spins which do not respond to a magnetic field.
This difference does not affect the velocity renormalization derived in this section, since the interaction is spin-independent. However, as we will show in the next section, this difference leads to the renormalization of the g-factor,  which occurs only in the surface state due to real spin Pauli matrices.

\section{g-factor renormalization}
In order to study the renormalization of the g-factor, we perform the WRG analysis for the Zeeman term. 
For simplicity, we consider the magnetic field to be perpendicular to the surface. Then the Zeeman Hamiltonian can be written as
\begin{equation}
\mathcal{H}_Z=\frac{g\mu_B}{2}B_z\sigma_z.
\end{equation}
Here $g$ and $\mu_B$ are the $g$-factor and the Bohr magneton, respectively.
The partition function for the surface state with the Zeeman term is given by
\begin{equation}
Z=\int \mathcal{D}(\bar{\Psi},\Psi)\mathrm{exp}[-S_0-S_{int}-S_{Z}], 
\end{equation}
where
\begin{equation}
S_{Z}=\frac{1}{\beta}\sum_{n}\int^{\Lambda} \frac{d\textbf{k}}{(2\pi)^2}\bar{\Psi}(\textbf{k},i\omega_n)(\frac{g\mu_B}{2}B_z \sigma_z)\Psi(\textbf{k},i\omega_n).
\end{equation}
Here we focus on the spin susceptibility and ignore the vector potential terms which contribute to the orbital susceptibility. 

The 1-loop correction to the  Zeeman term is represented by Fig.$\ref{fig1}$(b).
This diagram can be calculated as
\begin{align}
\delta \Gamma&=\frac{g\mu_B}{2}B_z(-T)\sum_n\sum_{\textbf{k}\in\delta\Lambda}\mathcal{G}(\textbf{k},i\omega_n)\sigma_z\mathcal{G}(\textbf{k},i\omega_n)\frac{2\pi e^2}{\epsilon |\textbf{k}|}\notag\\
&\cong\delta l \frac{ge^2}{2\epsilon v}\frac{\mu_B}{2}B_z \sigma_z.
\end{align} 
$\delta\Gamma$ changes the Zeeman action $S_Z$ and we obtain the effective Zeeman action
\begin{align}
S_{Z}'=\frac{1}{\beta'}\sum_{n}\int^{\Lambda} \frac{d\textbf{k}'}{(2\pi)^2}\bar{\Psi}'(\textbf{k}',i\omega_n')g(1+\delta l\frac{e^2}{2\epsilon v})\notag\\\frac{\mu_B}{2}B_z' \sigma_z\Psi'(\textbf{k}',i\omega_n'),\label{gfactor}
\end{align}
where
\begin{equation}
B_z'=B_z e^{\delta l}.
\end{equation}
Here we have performed the same rescaling as in the previous section. 
From this, we can construct the RG equation for the g-factor:
\begin{equation}
\frac{dg(l)}{dl}=g\frac{e^2}{2\epsilon}\frac{1}{v(l)}.\label{grg}
\end{equation}
Combining eqs.($\ref{velocity}$) and  ($\ref{grg}$), we obtain the effective g-factor
\begin{equation}
g(l)=g\left[1+2\mathrm{log}\left(1+\frac{e^2}{4v\epsilon}l \right)\right].
\end{equation}
In the temperature range of $T\ll T_0$, $g(T)$ can approximately be written as
\begin{equation}
g(T)\approx g\left[1+2\mathrm{log}\left[1+\frac{e^2}{4v\epsilon}\left(\mathrm{log}\frac{T_0}{T}\right) \right]\right].\label{gtemp}
\end{equation}
Eq.($\ref{gtemp}$) shows that the g-factor is enhanced by the LRCI at low temperatures similarly to the Fermi velocity except for the temperature dependence.

It is important to note that this g-factor renormalization is a unique characteristic of the surface state of a topological insulator.
To our knowledge, there have been no discussion about the renormalization group for the Zeeman term in Dirac electron systems. For graphene, the diagram in Fig.$\ref{fig1}$(b) has no contribution because Pauli matrices in the Green function do not represent real spins. 
Therefore the g-factor in graphene is a marginal coupling constant of the WRG analysis at 1-loop level.

\section{Spin susceptibility}
In this section, we calculate the spin susceptibility of the surface state of a topological insulator.

\begin{figure}[htbp]
\begin{center}
　　　\includegraphics[width=4.5cm,angle=90,clip]{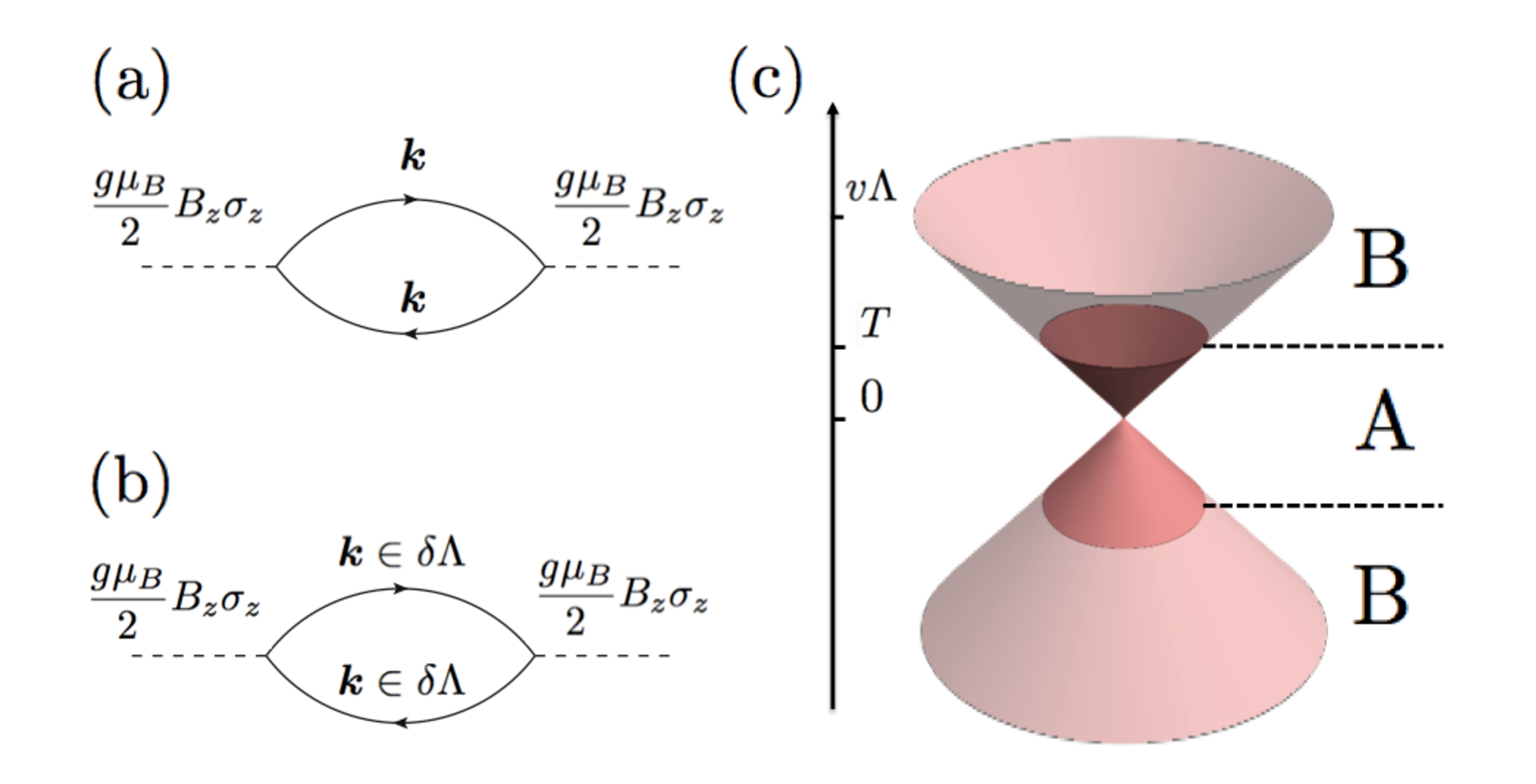}
　　　\caption{(a) Diagram of the free energy. (b) Diagram of the prefactor of the partition function (See Appendix). (c) Schematic picture of free energy contributions. A: contribution from renormalized electrons. B: contribution from the prefactor of the partition function. }
　　　\label{fig2}
\end{center}
\end{figure}

For the non-interacting case, the free energy corresponding to the spin susceptibility is represented by Fig.$\ref{fig2}$(a) and calculated as
\begin{align}
F&=\frac{g^2\mu_B^2B_z^2}{8}T\sum_n\sum_{\textbf{k}}Tr[\sigma_z\mathcal{G}(\textbf{k},i\omega_n)\sigma_z\mathcal{G}(\textbf{k},i\omega_n)]\notag\\
&=\frac{g^2\mu_B^2B_z^2}{8}\frac{1}{2\pi v^2}\int_{0}^{v\Lambda}d\epsilon G(\beta\epsilon)\equiv -\frac{1}{2}\chi_s^{(0)}B_z^2,\label{freecal}
\end{align}
where 
\begin{align}
G(t)&=f(t)-f(-t),\notag\\
f(t)&=\frac{1}{e^t+1}.
\end{align}
The spin susceptibility $\chi_s^{(0)}$ is obtained as 
\begin{equation}
\chi_s^{(0)}(T)=\frac{g^2\mu_B^2}{8}\frac{1}{\pi v^2}\left[v\Lambda+\frac{2}{\beta}\mathrm{log}\left(\frac{1+e^{-\beta v \Lambda}}{2}     \right)\right].\label{susceptibility}
\end{equation}

It is important to note that $\chi_s^{(0)}(T)$ depends on the cutoff $\Lambda$. 
This implies that the electrons apart from the Fermi energy contribute to the spin susceptibility.
In usual cases including graphene, the spin susceptibility is determined by the electrons within the temperature range near the Fermi energy$\cite{21}$.
In the present model, on the other hand, real spin Pauli matrices in the Dirac Hamiltonian give a non-trivial spin susceptibility.

In the following, we calculate the spin susceptibility with interaction. In the presence of the LRCI, we have shown that the Fermi velocity and the g-factor are enhanced. 
According to eq.($\ref{susceptibility}$), the enhancement of the g-factor increases the spin susceptibility, whereas that of the Fermi velocity decreases the spin susceptibility.  Therefore it is non-trivial whether or not the spin susceptibility is enhanced by the LRCI.

In this system, we find that there are two types of contributions to the spin susceptibility as shown in Fig. $\ref{fig2}$(c). One is the contribution from the renormalized electrons near the Fermi energy (A) and the other is from the prefactor of the partition function discussed later (B). 

\subsection{Contribution from renormalized electrons}
According to the WRG procedures in the previous sections, we obtain the effective quadratic actions
\begin{align}
S_0^{(l^*)}=\frac{1}{\beta(l^*)}\sum_{n}\int^{\Lambda}& \frac{d\textbf{k}}{(2\pi)^2}\bar{\Psi}^{(l^*)}\biggl[-i\omega_n(l^*)\textbf{1}\biggr.\notag\\
&\biggl. +v(l^*)(k_x\sigma_y-k_y\sigma_x)\biggr]\Psi^{(l^*)},\\
S_{Z}^{(l^*)}=\frac{1}{\beta(l^*)}\sum_{n}\int^{\Lambda} &\frac{d\textbf{k}}{(2\pi)^2}\bar{\Psi}^{(l^*)}\frac{g(l^*)\mu_B}{2}B_z \sigma_z\Psi^{(l^*)}.
\end{align}
These effective actions contain the renormalized electrons within the temperature range near the Fermi energy. At low energy scale, α is sufficiently small and we can approximate the free energy as
\begin{align}
F_s^<(T(l^*))&=\frac{\mu_B^2B_z^2e^{2l^*}}{8}\frac{g(l^*)^2}{2\pi v(l^*)^2}\int_{0}^{v(l^*)\Lambda}d\epsilon G(\beta(l)\epsilon)\notag\\
&=\frac{\mu_B^2B_z^2e^{3l^*}}{8}\frac{g(l^*)^2T}{2\pi v(l^*)^2}\int_{0}^{1}dtG(t),\label{fs<*}
\end{align}
as in the non-interacting case.
Here we have used eq.($\ref{tvrelation}$) in the second line. Scaling law for the free energy can be written as
\begin{align}
F(T)=e^{-2l^*}\frac{T}{T(l^*)}F(T(l^*)).\label{scaling}
\end{align}
Combining eqs.($\ref{fs<*}$) and  ($\ref{scaling}$), we obtain the free energy before rescaling  
\begin{equation}
F_s^<(T)=\frac{\mu_B^2B_z^2}{8}\frac{g(l^*)^2T}{2\pi v(l^*)^2}\int_{0}^{1}dtG(t),\\
\end{equation}
and the spin susceptibility from the renormalized electrons can be calculated as
\begin{equation}
\chi_s^<(T)=\mathrm{log}\left[\frac{(1+e)^2}{4e}\right]\frac{\mu_B^2g(l^*)^2T}{8\pi v(l^*)^2}.\label{chi<}
\end{equation}
\subsection{Contribution from the prefactor of the partition function}
In the above procedure, we have focused only on the action and neglected the prefactor of the partition function like $Const.$ in eq.($\ref{Const}$).
This is because we only need the effective action to obtain the RG equation.
To obtain the free energy, however, we need the whole partition function including the prefactor.
In Appendix, we show that this contribution is represented by Fig. $\ref{fig2}$(b).
In the $l\rightarrow l+\delta l$ step, this diagram can be calculated as (Appendix)
\begin{align}
\delta F_s^>(T(l))&=\frac{\mu_B^2B_z^2e^{2l}}{8}\frac{g(l)^2}{2\pi v(l)}\int_{\Lambda e^{-\delta l}}^{\Lambda}dkG(\beta(l)v(l)k)\notag\\
&=\frac{\mu_B^2B_z^2e^{3l}}{8}\frac{g(l)^2}{2\pi v(l)}\int_{\Lambda e^{-(l+\delta l})}^{\Lambda e^{-l}}dkG(\beta v(l)k).\label{deltaf}
\end{align}
Using the scaling law $(\ref{scaling})$ and the following relation
\begin{equation}
k\cong\Lambda e^{-l},
\end{equation}
we obtain the contribution 
\begin{equation}
\delta F_s^>(T)=\frac{\mu_B^2B_z^2}{16\pi }\int_{\Lambda e^{-(l+\delta l)}}^{\Lambda e^{-l}}dk\frac{g(k)^2}{v(k)}G(\beta v(k)k),\label{delta}
\end{equation}
where
\begin{align}
v(k)&=v+\frac{e^2}{4\epsilon}\mathrm{log}\frac{\Lambda}{k},\\
g(k)&=g\left[1+2\mathrm{log}\left[1+\frac{e^2}{4v\epsilon}\mathrm{log}\frac{\Lambda}{k}\right]\right].
\end{align}
The whole contribution apart from the Fermi energy can be calculated by summing up eq.($\ref{delta}$) in $l=0\rightarrow l^*$: 
\begin{equation}
F_s^>(T)=\frac{\mu_B^2B_z^2}{16\pi }\int_{\Lambda e^{-l^*}}^{\Lambda}dk\frac{g(k)^2}{v(k)}G(\beta v(k)k),
\end{equation}
and the corresponding spin susceptibility is 
\begin{equation}
\chi_s^>(T)=\frac{-\mu_B^2}{8\pi }\int_{\Lambda e^{-l^*}}^{\Lambda}dk\frac{g(k)^2}{v(k)}G(\beta v(k)k).\label{chi>}
\end{equation}
\subsection{total spin susceptibility}
Combining eqs.($\ref{chi<}$) and  ($\ref{chi>}$), we obtain the total spin susceptibility
\begin{align}
\chi_s(T)=&\mathrm{log}\left[\frac{(1+e)^2}{4e}\right]\frac{\mu_B^2g(l^*)^2T}{8\pi v(l^*)^2}\notag\\
-&\frac{\mu_B^2}{8\pi }\int_{\Lambda e^{-l^*}}^{\Lambda}dk\frac{g(k)^2}{v(k)}G(\beta v(k)k).\label{totalsusceptibility}
\end{align}

\begin{figure}[htbp]
\begin{center}
　　　\includegraphics[width=8cm,angle=0,clip]{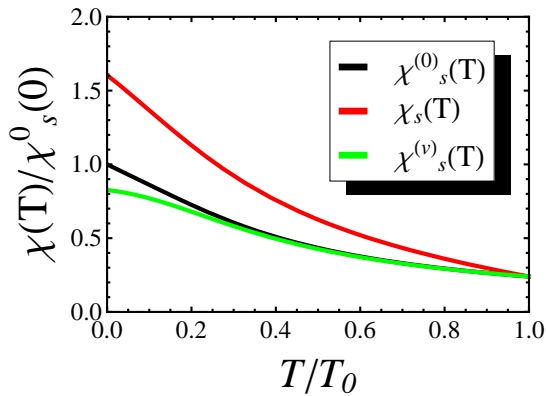}
　　　\caption{Spin susceptibility vs temperature. $\chi_s^{(0)}(0)$ is the spin susceptibility without interaction at T=0. $T_0$ is defined as the cutoff energy $v\Lambda$. The magnitude of the interaction $\alpha$ is 1. Black line: Non-interacting spin susceptibility $\chi_s^{(0)}(T)$ given by eq.($\ref{susceptibility}$). Red line: Total spin susceptibility $\chi_s(T)$ given by eq.($\ref{totalsusceptibility}$). Green line: $\chi_s^{(v)}(T)$ given by eq.($\ref{vsus}$).  }
　　　\label{fig3}
\end{center}
\end{figure}

We can not conclude at first sight whether or not the total spin susceptibility is enhanced by the LRCI because the second term can not be calculated analytically. 
Therefore we performed numerical integration. The result of numerical integration in the case of $\alpha=1$ is shown in Fig.$\ref{fig3}$. For the sake of comparison, we show the non-interacting spin susceptibility $\chi_s^{(0)}(T)$ given by eq.($\ref{susceptibility}$).
In order to clarify the importance of the g-factor renormalization, we also show the spin susceptibility, which only includes the Fermi velocity renormalization
\begin{align}
\chi_s^{(v)}(T)=&\mathrm{log}\left[\frac{(1+e)^2}{4e}\right]\frac{\mu_B^2g^2T}{8\pi v(l^*)^2}\notag\\
-&\frac{\mu_B^2}{8\pi }\int_{\Lambda e^{-l^*}}^{\Lambda}dk\frac{g^2}{v(k)}G(\beta v(k)k).\label{vsus}
\end{align}
Here we have replaced the renormalized g-factor with the bare one in eq.($\ref{totalsusceptibility}$).
Fig.$\ref{fig3}$ shows that the effect of the g-factor renormalization overcomes that of the Fermi velocity renormalization, and the spin susceptibility is enhanced at low temperature.

\section{summary}
In this paper, we have investigated the Dirac electrons in the surface state of a three-dimensional topological insulator in the presence of the LRCI.
Using the Wilsonian renormalization group, we have found two renormalization effects. One is the Fermi velocity renormalization, which occurs in a usual Dirac system. The other is the g-factor renormalization, which occurs only in our Dirac system. The latter effect
appears since Pauli matrices in the surface Dirac Hamiltonian represents real spins.
We have also calculated the spin susceptibility of our system. Numerical integration has shown that the g-factor renormalization enhances the spin susceptibility, overcoming the reducing effect which comes from the Fermi velocity renormalization. In a usual Dirac system, on the other hand, the spin susceptibility is suppressed by the LRCI$\cite{21}$ due to the lack of the g-factor renormalization.
Our work has clarified the importance of the LRCI in spin response when a Dirac Hamiltonian contains real spin Pauli matrices.

\begin{acknowledgements}
This work was supported by Japan Society for the Promotion of Science through Program for Leading Graduate Schools (MERIT) and Grants-in-Aid for Scientific Research (A) (No. 24244053).
\end{acknowledgements}
\appendix*
\section{Spin susceptibility contribution from the prefactor of the partition function}
In this appendix, we show that the prefactor of the partition function, which are not used in the RG equations, contribute to the spin susceptibility represented by Fig. $\ref{fig2}$(b).

In the $l\rightarrow l+\delta l$ step, the partition function is deformed as
\begin{align}
Z=& Const. \int \mathcal{D}(\bar{\Psi},\Psi)\mathrm{exp}\left[-S_0^{(l)}-S_Z^{(l)}-S_{int}\right]\notag\\
=&Const. \int \mathcal{D}(\bar{\Psi}_>,\Psi_>)\mathrm{exp}\left[-S_0^{(l)}-S_Z^{(l)}\right]\times\notag\\
&\Biggl\{   \frac{\int \mathcal{D}(\bar{\Psi}_>,\Psi_>)\mathrm{exp}\left[-S_0^{(l)}-S_Z^{(l)}-S_{int}\right]}{\int \mathcal{D}(\bar{\Psi}_>,\Psi_>)\mathrm{exp}\left[-S_0^{(l)}-S_Z^{(l)}\right]}\times\Biggr.    \notag\\
&\Biggl.\int \mathcal{D}(\bar{\Psi}_<,\Psi_<)\mathrm{exp}\left[-S_0^{(l)}-S_Z^{(l)}\right]\Biggr\}.
\end{align}
The factors in the latter parenthes are used in the RG equations, while the first factor is not needed. However, in the calculation of the spin susceptibility, the first factor contributes to the spin susceptibility.
The first factor in the $l\rightarrow l+\delta l$ step can be rewritten as
\begin{align}
&\int \mathcal{D}(\bar{\Psi}_>,\Psi_>)\mathrm{exp}\left[-S_0^{(l)}-S_Z^{(l)}\right]\notag\\
&=\mathrm{exp}\left[\mathrm{ln}\ \mathrm{det}_{\textbf{k}\in\delta\Lambda,\omega_n}\left(-{\mathcal{G}^{(l)}}^{-1}+\mathcal{H}_Z^{(l)}   \right)   \right].
 \end{align}
Using $\mathrm{ln}\ \mathrm{det}=\mathrm{tr}\ \mathrm{ln}$, we can see that this factor gives the contribution to the free energy as
\begin{align}
\delta F^{(l)}&=-\mathrm{ln}\ \mathrm{det}_{\textbf{k}\in\delta\Lambda,\omega_n}\left(-{\mathcal{G}^{(l)}}^{-1}+\mathcal{H}_Z^{(l)}   \right)\notag\\
&=- \mathrm{tr}_{\bm{k}\in\delta\Lambda,\omega_n}\ \mathrm{ln}\left(-{\mathcal{G}^{(l)}}^{-1}+\mathcal{H}_Z^{(l)}   \right) \notag\\
&=-\mathrm{tr}_{\bm{k}\in\delta\Lambda,\omega_n}\ \mathrm{ln}\left(-{\mathcal{G}^{(l)}}^{-1}\right)-\mathrm{tr}\ \mathrm{ln}\left(1-\mathcal{G}^{(l)}\mathcal{H}_Z^{(l)}\right)\notag\\
&\cong\frac{1}{2}\mathrm{tr}_{\bm{k}\in\delta\Lambda,\omega_n}\left[\mathcal{G}^{(l)}\mathcal{H}_Z^{(l)}\mathcal{G}^{(l)}\mathcal{H}_Z^{(l)}\right].\label{shoumei}
\end{align}
Here we only keep the 2nd order contribution of the magnetic field. Eq. ($\ref{shoumei}$) is calculated as in eq.($\ref{freecal}$) and we finally obtain eq.($\ref{deltaf}$).

\end{document}